\documentclass{IEEEtran}
\usepackage{multirow}

\usepackage{cite}
\usepackage{amsmath,amssymb,amsfonts}
\usepackage{algorithmic}
\usepackage{graphicx}
\usepackage{textcomp}
\usepackage[table]{xcolor}
\usepackage{soul}
\newtheorem{defn}{Definition}
\def\BibTeX{{\rm B\kern-.05em{\sc i\kern-.025em b}\kern-.08em
		T\kern-.1667em\lower.7ex\hbox{E}\kern-.125emX}}

\usepackage{caption,setspace}

\begin{document}
	\title{Risk, Trust, and Bias: Causal Regulators of Biometric-Enabled Decision Support}
	
	\author{\IEEEauthorblockN{ 
			Kenneth Lai$^{1}$, Helder C. R. Oliveira$^{1}$, Ming Hou$^{2}$, Svetlana N. Yanushkevich$^{1}$, and Vlad Shmerko$^{1}$
		}\\
		\IEEEauthorblockA{
			\textit{$^1$Biometric Technologies Laboratory, Department of ECE,} \textit{University of Calgary, Canada,}\\
			Web: http://www.ucalgary.ca/btlab, E-mail: \{kelai,helder.rodriguesdeol,syanshk,vshmerko\}@ucalgary.ca\\
		\textit{$^2$Defence Research and Development Canada (DRDC), Canada,}\\
		 E-mail: ming.hou@drdc-rddc.gc.ca
	}}
		\maketitle
		\thispagestyle{empty}
	\begin{abstract}
		Biometrics and biometric-enabled decision support systems (DSS) have become a mandatory part of complex dynamic systems such as security checkpoints, personal health monitoring systems, autonomous robots, and epidemiological surveillance. Risk, trust, and bias (R-T-B) are emerging measures of performance of such systems. The existing studies on the R-T-B impact on system performance mostly ignore the complementary nature of R-T-B and their causal relationships, for instance, risk of trust, risk of bias, and risk of trust over biases. This paper offers a complete taxonomy of the R-T-B causal performance regulators for the biometric-enabled DSS. The proposed novel taxonomy links the R-T-B assessment to the causal inference mechanism for reasoning in decision making. \textbf{Practical details} of the R-T-B assessment in the DSS are demonstrated using the experiments of assessing the trust in synthetic biometric and the risk of bias in face biometrics. The paper also outlines the \textbf{emerging applications} of the proposed approach beyond biometrics, including decision support for epidemiological surveillance such as for COVID-19 pandemics.
		
	\end{abstract}
	
	

	\section{Introduction}
	\label{sec:}
	\IEEEPARstart{B}{}iometric-enabled decision support is a mandatory mechanism of various complex systems, such as: 
	\begin{itemize}
		\item [$-$]security checkpoints (identity management) \cite{[Labati-2016]},
		\item [$-$]personal health monitoring systems \cite{[Andreu-2016]},
		\item [$-$]e-coaching for health \cite{[Ochoa-2018]},
		\item [$-$]driver assistant (e.g., fatigue and stress detection) \cite{[Rigas-2012]},
		\item [$-$]multi-factor authentication systems \cite{[Roy-2018]},
		\item [$-$]spoken conversational agents \cite{[Weng-2016]}, 
		\item [$-$]epidemiological surveillance \cite{[McLachlan-2020]} and,
		\item [$-$]preparedness systems for emerging health service \cite{[Neville-2015]}.
	\end{itemize}
	
	To be integrated into a complex system, a biometric-enabled computational intelligence (CI) must satisfy various requirements for compatibility and standards. In other words, it must adhere to the concept of Decision Support System (DSS). Generally, the goal of any DSS is to support human experts, operators, or users in their decision-making in real-time, under multiple and constantly evolving factors. 
	
	A well-identified trend in DSS is to augment \textbf{intelligent features} (learning, training, adaptation, possibility to choose among available decisions) beyond simple information retrieving, e.g., predicting the evolution of the current state situation, known as situational awareness \cite{[Poursaberi-2013]}. In intelligent DSS, human cognition can impact the CI, and vice versa. That is, the performance of the DSS depends on complicated factors such as cognitive biases of humans and intelligent machines. This is the area of our interest.
	
	Typically, the DSS performance is evaluated in various dimensions:
	\begin{itemize}
		\item [$-$]Technical, e.g., false acceptance rate (FAR), false rejection rate (FRR), accuracy rate, and throughput \cite{[Roy-2018],Bolle},
		\item [$-$]Social, e.g., privacy, public acceptance \cite{[Andreou-2017]},
		\item [$-$]Psychological, e.g., efficiency of human-machine interactions (known as teaming and trustworthy intelligent systems) \cite{[Hu-2019],[Hugenberg-2013],[Montibeller-2015]}, 
		\item [$-$]Security, e.g., vulnerability and sensitivity of personal data 
		\cite{[Andreou-2017],[Montibeller-2015],[Yanush-2019]}, and
		\item [$-$]Efficiency of teamwork and group decision, e.g., trust, risk, reliance, satisfaction, stress \cite{[Brown-2020]}.
	\end{itemize}
	
\begin{figure}[!ht]
		\begin{center}
			\includegraphics[scale=0.65]{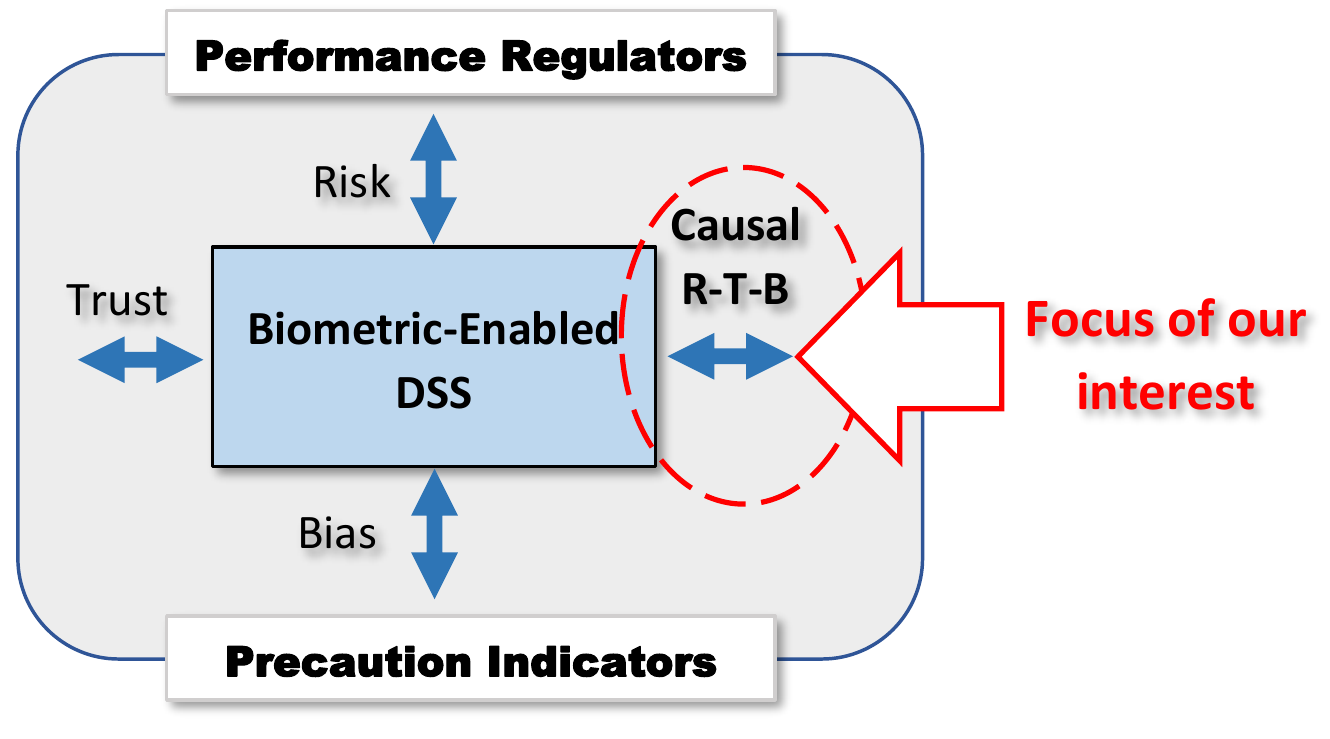}
		\end{center}
		\caption{The R-T-B impact on the DSS performance and can be used as precaution indicators. The complementary nature of the R-T-B is the focus of our study.}
		\label{fig:Risk-Trust_Bias_Regulators}
	\end{figure}

	\emph{Risk, Trust}, and \emph{Bias} (R-T-B) are essential indicators of the performance evaluation of complex dynamic systems such as intelligent DSS (Fig. \ref{fig:Risk-Trust_Bias_Regulators}). The R-T-B measures belong to the class of high-level performance measures. For example, {trust} in the intelligent interview assistant addresses the intelligent (cognitive) biases \cite{[Yanush-2019]}. {Risk} and {trust} in the DSS are linked to various kinds of biases, for example, racial biases in face-based human identification \cite{[Das-2019],[Grother-2019]} and attribute biases in social profiles \cite{[Andreou-2017]}. 
	
	As follows from Fig. \ref{fig:Risk-Trust_Bias_Regulators}, the role of the R-T-B measures is twofold: \emph{performance regulators} and \emph{precaution indicators}. Contemporary approaches often consider the R-T-B impact on the DSS performance \textbf{independently, ignoring their causal relationships} such as the risk of bias, the risk of trust, the risk of trust over biases, etc. Adding to the evaluation level, the R-T-B measures become precaution indicators or signs, e.g., high risk, low trust, large bias, and low risk of bias.
	
	\textbf{The complementary nature of the R-T-B measures in biometric-enabled DSS is the focus of our study} (Fig. \ref{fig:Risk-Trust_Bias_Regulators}). Specifically, R-T-B can manifest themselves as a \emph{causal ensemble} and convey additional useful information for DSS performance regulation. {For example, once a decision regarding a subject's identity is made, a ``risk of decision trust'' is calculated that  1) assesses the risk of acceptance or rejection of the decision based on the operator's trust in the DSS, and 2) acts as a measure of precaution on the over-trust. The \textbf{working hypothesis} of our research is that the R-T-B landscape that includes the causality between the R-T-B measures is evaluated using causal networks.} In this paper, we provide the results of such study. 
	
	The remainder of this paper is organized as follows. Section \ref{sec:Related-work} provides a survey of the most important related works. Contributions of this paper are provided in Section \ref{sec:Motivation}. Our approach to the R-T-B taxonomy is explained in Sections \ref{sec:Background}, \ref{sec:Fundamental-operations}, including the view on standardization of the R-T-B measures using the Admiralty Code is proposed in Section \ref{sec:Admiralty-Code}. {The core mechanism of the R-T-B assessment on the causal networks is explained in Section \ref{sec:Syst-view-causal-nets} and demonstrated through experiments in Section \ref{sec:Exp}}. The forecast of emerging applications and overall summary are provided in Sections \ref{sec:Special-Section} and \ref{sec:Summary_conclusions}.
	
	\section{Related work}
	\label{sec:Related-work}
	
\begin{figure}[!ht]
		\begin{center}
			\includegraphics[scale=0.65]{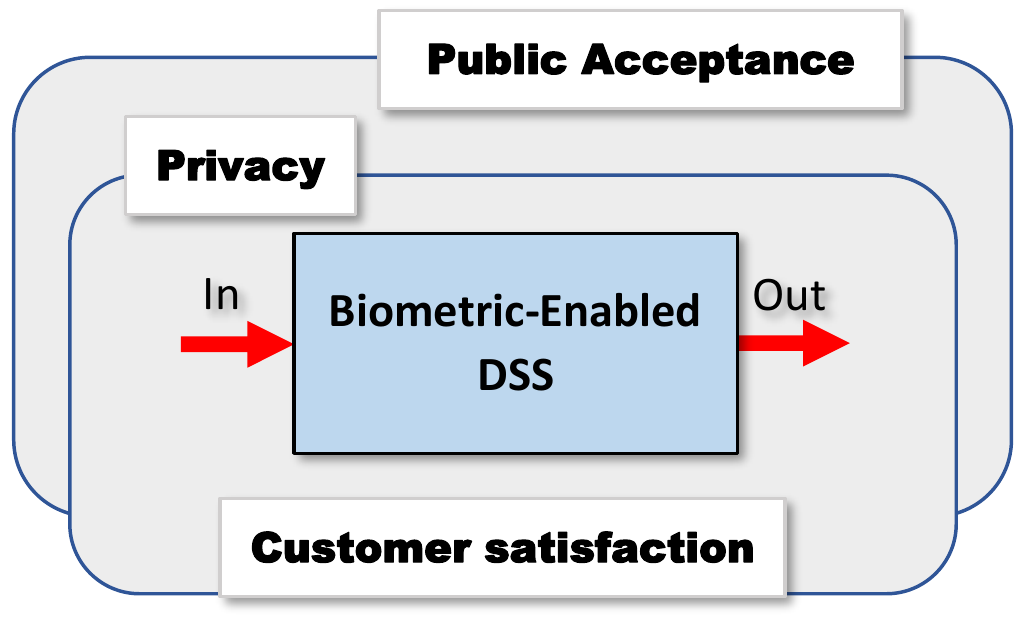}
		\end{center}
		\caption{Typical integrated performance measures of the biometric-enabled DSS are privacy and public acceptance.}
		\label{fig:Measures-Related}
	\end{figure}

	Performance measurement is commonly understood as a regular measurement of a system's outcomes that captures the efficiency of said system. Measures such as privacy, customer satisfaction, and public acceptance belong to a class of \emph{integrated, or high-level} system performance measures (Fig. \ref{fig:Measures-Related}). Each of these measures includes a specific set of quantitative and qualitative indicators, often complementary. For example, 
	\begin{itemize}
		\item [$-$]\emph{privacy} includes indicators such as personal data (collection, storing, sharing, etc.); 
		\item [$-$]\emph{public acceptance} includes security, privacy, user satisfaction, etc. 
		\item [$-$]\emph{common indicators} of privacy and public acceptance measures include psychological predictors, social profile factors, and demographic indicators.
	\end{itemize}
	
	In order to achieve the compatibility between systems placed on a unified computational platform, it is reasonable to extend these measures using the R-T-B indicators like ``risk of storing personal data'', ``demographic bias'', and ``trust of social profiling''. Some of the R-T-B projections have been studied in the last decade in a wide spectrum of applications, for example, disclosure risks \cite{[Andreou-2017]} (privacy losses in social computing networks), biases in facial recognition \cite{[Das-2019]}, trust, risk and optimism bias in e-government \cite{[Schaupp-2010]}, risk of crime and social trust in the presence of endogeneity bias \cite{[Zanin-2013]}. In \cite{[Zhou-2013]}, the notion of `biased trust' addresses the phenomenon of a small set of trusted network users. The adversary can use this bias (prior trust relationships) for the development of an attack strategy in onion-routing networks. In \cite{[Lee-2014]}, risks of decision-making biases and biases of the trustworthiness are studied for various consumer scenarios. In \cite{[Gross-2016]}, the trust of reduced risk has been used for mobile shopping analysis.
	
	The risk of bias in CI judgments is a key interest in all CI applications, e.g., in medicine \cite{[Gates-2018]} (trust of machine decision) and security \cite{[Grother-2019]} (risk of mis-identification due to ``demographic'' bias in facial recognition algorithms). Assessing the trust of these phenomena for a given and novel CI algorithm is of critical importance.	Inappropriate calibration of trust in human-machine and machine-machine interactions is a serious problem, and when conjoined with bias, risks of various kind of unwanted effects greatly intensifies. Among various approaches, intelligent DSSs are of the greatest demand, e.g., risk-adaptive trust model \cite{[Van-hamme-2018]}. 
	
	
	Paper \cite{[Lopez-2015]} is an introduction to the trust management engine using pattern recognition techniques. The key notion is the \emph{trust feature}: ``the desired feature to be taken into consideration for a trust assessment''. Example of established trust features include knowledge, reputation, and experience. \emph{Trust assessment classes} include ``untrustworthy'', ``neutrally trusted'', and ``trustworthy''. Next, the regulators (measures) of trust discipline are the \emph{trust levels}. The inputs of the machine learning algorithms are the trust features in a certain context, $\mathbf{X}=\texttt{Context}$, some labels $y^{(i)}, i=1,2,\ldots , m$, are assigned to each training set, $\{\textbf{X},y^{(i)} \}$. This is a formulation of a trust assessment problem in terms of pattern recognition. In particular, various opportunities for choosing an appropriate machine learning algorithm are provided, e.g., multi-class classifiers such as a deep learning network, or a support vector machine that trains the model in order to identify the best margin to separate the trustworthy interactions from the other interactions. More details on trust computing using machine learning algorithms are provided in \cite{[Jayasinghe-2019]}. In our opinion, the approach proposed in \cite{[Lopez-2015],[Jayasinghe-2019]} can be extended to the broader R-T-B spectrum.
	
	The above review leads to the conclusion that neither the R-T-B measures nor their causal relationships have been systematically addressed so far. Our study aims at overcoming these gaps and introduce the state-at-the-art R-T-B taxonomy and related cause-and-effects. 
	
	\section{Contributions}
	\label{sec:Motivation}
	
	The foundation for the proposed taxonomy was laid in \cite{[Lai-2020]} and \cite{[Yanush-2020-b]}. In \cite{[Lai-2020]}, risks of biases for facial recognition were investigated, and in \cite{[Yanush-2020-b]}, risk and trust indicators of synthetic data in cognitive security checkpoints were studied. The quintessence of the experimental results from these sources is analyzed in Section \ref{sec:Exp}. This paper takes these results one step further by introducing a \textbf{systematic approach} to the R-T-B causal performance evaluation of complex biometric-enabled systems. 
	
	\textbf{Our contribution and goal are achieved in conjunction with the following results:} 
	
	\begin{itemize}
		\item [$-$] \textbf{Framework of intelligent DSS}; we adopted \emph{Haykin's} fundamental results on cognitive systems \cite{[Haykin-2012]}, in particular, in modeling the DSS for multi-state intelligent checkpoint;
		\item [$-$] \textbf{Taxonomical view} for causal R-T-B inference; we adopted, for this purpose, \emph{Pearl's} layered causal inference hierarchy \cite{[Pearl-2019]}, as well as fundamentals of causal (Bayesian) networks \cite{[Pearl-2000]};
		\item [$-$] \textbf{Standardization the R-T-B measures}; we referred to a widely used in practice the Admiralty Code \cite{[Admiralty_Code_2012],[U.S._Army_Field_Manual],[Blasch-2013]}; and
		\item [$-$] \textbf{Systematic view} of advanced causal networks; we extended a recent review \cite{[Rohmer-2020]} that covers most of the causal networks.
	\end{itemize}

	\section{Background}
	\label{sec:Background}
	
	This Section provides the basic knowledge of the intelligent DSS over the R-T-B performance regulators. 
	
	\subsection{Framework of intelligent DSS}
	
\begin{figure}[!ht]
		\begin{center}
			\includegraphics[scale=0.65]{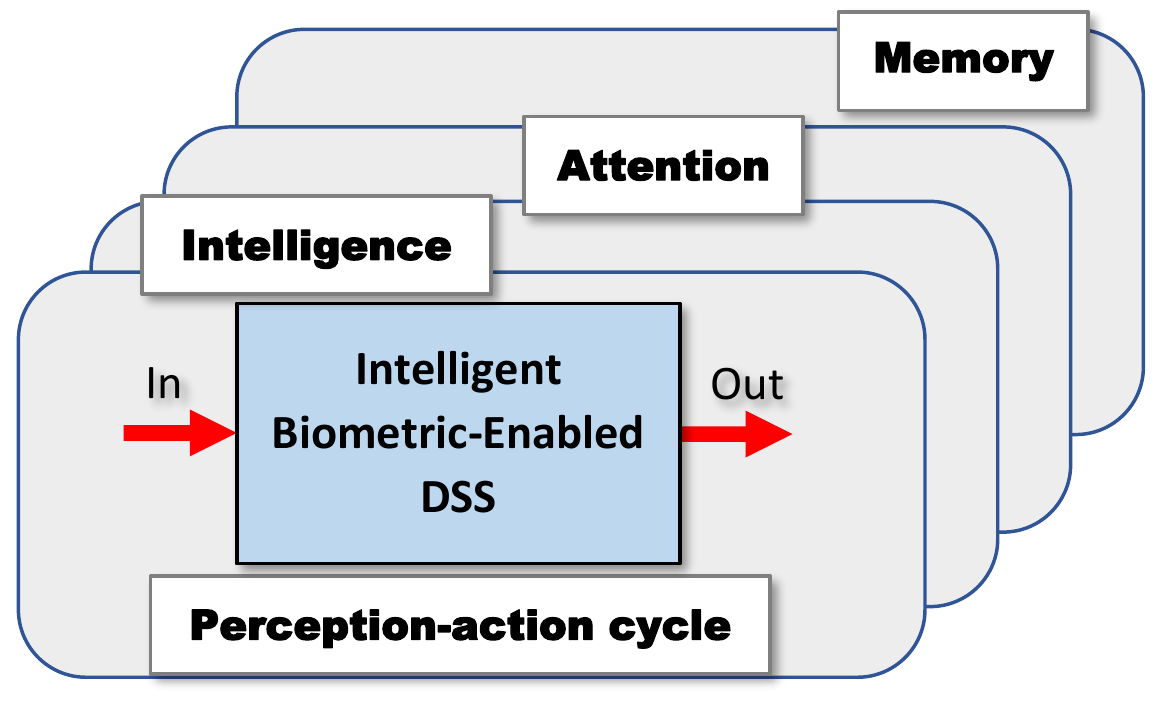}
		\end{center}
		\caption{Principle elements of the intelligent DSS.}
		\label{fig:Intelligent-DSS}
	\end{figure}

	Intelligent biometric-enabled DSS for identity management is a complex dynamic system \cite{[Yanush-2019],[Yanush-2019-a]} with the following elements of a {cognitive} system \cite{[Haykin-2012]} (Fig. \ref{fig:Intelligent-DSS}):
	\begin{enumerate}
		\item []\hspace{-9mm} \emph{Perception-action cycle} that enables information gain regarding the state of an identified person;
		\item []\hspace{-9mm} \emph{Memory} distributed across the entire system (personal data are collected in the physical and virtual world); 
		\item []\hspace{-9mm} \emph{Attention} is driven by memory to prioritize the allocation of available resources; and 
		\item []\hspace{-9mm} \emph{Intelligence} is driven by perception, memories, and attention; its function is to enable the control and decision-making mechanism to help identify intelligent choices. 
	\end{enumerate}
	
	The R-T-B measures are an integrated part of the high-level measurements used in intelligent systems, e.g., risk of person identification in the perception-action cycle, privacy trust of distributed memory, and attention bias. Intelligent DSS is a semi-automated system, which deploys CI to process the data sources and to assess R-T-B; this assessment is submitted to a human operator for the final decision.
	
	\subsection{The R-T-B definitions}
	
	\begin{defn}
		\textbf{Risk} is a ``measure of the extent to which an entity is threatened by a potential circumstance or event, and typically is a function of: (i) the adverse \textbf{impact}, also called \textbf{cost} or magnitude of harm, that would arise if the circumstance or event occurs, and (ii) the \textbf{likelihood} of event occurrence''~\cite{[NIST-2017]}. $\blacksquare$
	\end{defn}

	Formally, \texttt{risk} in this paper is defined as a function $F$ of impact (or consequences) of a circumstance or event and its occurrence probability: 
	\[\texttt{Risk} = F(\texttt{Impact},\ \texttt{Probability})\]
	For example, in automated decision making, the \texttt{Risk} is expressed as an \texttt{Impact} of accepting the DSS decision (which can be correct or incorrect) magnified by the likelihood of its correctness or incorrectness. 
	
	For computational purpose of this paper, we adapt the following definition of trust.
	
	\begin{defn}
		\textbf{Trust} as defined in \cite{[Gambetta-2000],[Solhaug-2007]} is ``the \emph{subjective probability} by which one entity (the trustor) expects that another entity (the trustee) to perform a specific action of which its goal is dependent on.'' ~$\blacksquare$
	\end{defn}

	Useful details are provided in \cite{[Lee-2006]}: ``trust is the attitude that an agent (DSS in our case) will help achieve an individual's goals in a situation characterized by uncertainty and vulnerability.''
	
	From these definitions follows the key property of trust: the level of trust is \emph{believed probability}.
	
	The framework of trust assessment includes the following issues \cite{[Lee-2006],[Van-hamme-2018],	[Hu-2019],[Cho-2015]}:
	
	\begin{itemize}
		\item [$-$]The concept of trust is subjective by nature, its definition depends on a particular area.
		\item [$-$]In order to assess trust, it is reasonable to derive trust as a function of risk.
		\item [$-$]Trust is not necessarily proportional to the inverse of risk because risk may exist even under a situation with high trust.
		\item [$-$]{The balance between trust and risk can be achieved by optimizing gains in decisions.}
		\item [$-$]{In the presence of uncertainty, trust can provide a ``credit'' to decisions made under uncertainty.}
	\end{itemize}
	
	In \cite{Cohen97}, trust is associated with the expected utility of the decision, expressed in terms of \emph{cost of verification} and \emph{cost of the determined action}. Trust in the currently preferred decision on the action $a$ is the probability $P(a)$ that $a$ is successful. If trust is high enough, the currently preferred action can be accepted without verification; however, the operator will take risks for not performing verification. For example, face matching (as binary decision) by an operator aided with the DSS has only two options: Acceptance of the DSS recommendation, or its Rejection. If the option Accept is chosen, then the Trust in the DSS is simply the probability that the Accept decision is correct
	\[\texttt{Trust} = P(\texttt{Accept}) \]
	and the user should accept the DSS recommendation without verification if $\texttt{Trust} > 1 - C_v / C_a$ where $C_v$ is the cost of decision verification action and $C_a$ is the cost of incorrectly accepting the DSS solution. 
	
	\begin{defn}
		\textbf{Trustworthiness} as defined in \cite{[Gambetta-2000],[Solhaug-2007]} is ``the \emph{objective probability} by which the trustee performs a given action on which the goal of trustor depends''.~$\blacksquare$
	\end{defn}
	Useful details are given in \cite{[NIST-2017]}:	``trustworthiness is the degree to which a system can be expected to preserve the confidentiality, integrity, and availability of the information being processed, stored, or transmitted by the system across the full range of threats.''
	
	Hence, in contrast to trust, trustworthiness refers to the actual probability by which the trusted party will perform as expected.
	
	Note that trust is a belief that does not necessarily require observed behavior in the past, that is distinct from trustworthiness, which is a verified objective of trust through observed evidence. For example, a trusted biometric sample acquisition system should satisfy a set of requirements such as resistance to: 1) fake biometric target presentation, 2) communication attack, and 3) acquisition system tampering. In \cite{[Roy-2018]}, trustworthy values are calculated using the accuracy rates (like FAR and FRR) of different authentication modalities, that is, the higher accuracy rate makes the	modality more reliable.
	
	\begin{defn}
		\textbf{Bias} in the cognitive DSS refers to the tendency of an assessment process to systematically over- or under-estimate the value of a population parameter.~$\blacksquare$
	\end{defn}
	For example, the bias of trust refers to a phenomenon that is well identified in psychology. In our study, we consider the bias of trust as the difference between the baseline trust ($\texttt{Trust}$) and the trust given some prior knowledge on a specific parameter of the system ($\texttt{Trust}'$):
	\[\texttt{Bias}_{\texttt{Trust}} = \texttt{Trust} - \texttt{Trust}' \]
	The result can be positive (decreased trust) or negative (increased trust).
	
	Identifying and mitigating bias is essential for building trust and estimating the risk of trust in human-machine and machine-machine interactions. For example, the phenomenon of \emph{own-race bias} is well-known in psychology. It is a tendency for systems to better recognize faces from one's racial in-group rather than for racial out-groups \cite{[Hugenberg-2013]}. This tendency was recently shown in face recognition experiments \cite{[Das-2019],[Grother-2019],[Merler-2019]}. CI biases were introduced by intelligent support of human-machine interactions \cite{[Yanush-2019]}. Identity management biases were analyzed in multiple social profiles \cite{[Andreou-2017]}. In \cite{[Roy-2018]}, fuzzy reasoning and intelligent adaptive selection of the Multi-Factor Authentication (MFA) credentials have been used. Trustworthiness value functions are used in different authentication modalities such as biometrics, non-biometrics, and cognitive behavior metric. Fuzzy DSS for MFA is a cognitive system where a decision regarding user authentication is made iteratively and adaptively.

	\section{Taxonomical view on the causal R-T-B inference}
	\label{sec:Fundamental-operations}
	
	This section represents a crucial part of our approach. The complementary nature of the R-T-B triad is a well-known fact for researchers. Some of the R-T-B causal relationships are successfully used in various fields (see our review of related work in Section \ref{sec:Related-work}). Gaps between the R-T-B complementary nature and computational methodologies are periodically reviewed \cite{[Cho-2015],[EU-Fundamental-Rights-2019]}, standardized \cite{[ISO-31000-2009]} and created into guidelines \cite{[NIST-2017]}. The focus of our interest is motivated by the improvement of the DSS performance using machine learning techniques. 
	
\begin{figure}[!ht]
		\begin{center}
			\includegraphics[scale=0.5]{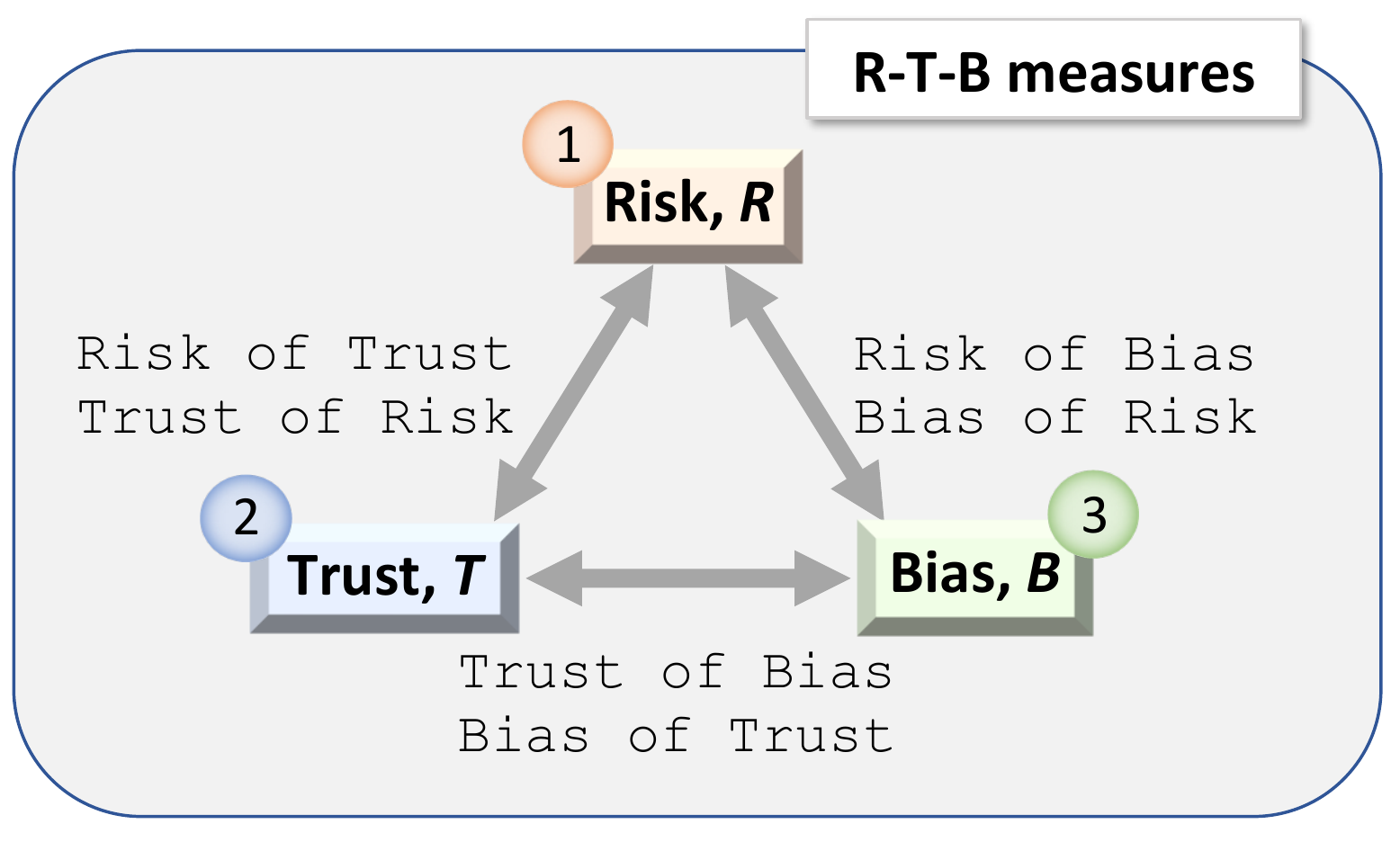}
		\end{center}
		\caption{The R-T-B causal landscape. }
		\label{fig:RiskTrustBias}
	\end{figure}

	There are two questions we address: 1) the reflection of the complementary nature of the R-T-B and 2) the need to close the aforementioned gaps. The principal solution to the first question is illustrated in Fig. \ref{fig:RiskTrustBias}. This R-T-B ensemble can be ordered according to their appropriate discipline. For example, first-order complement quantifies only single factor of the R-T-B domain and is introduced by a single node (variable), e.g., risk $R$ of event $X=x$. There are also second and third order R-T-B complements. 
	
	The next step is more complicated and addresses the nature of causal models \cite{[Pearl-2000]}. Recent work by \emph{Pearl} \cite{[Pearl-2019]} is a structured view on such model types. 
	
	Table \ref{tab:Causal-inference-hierarchy} is \emph{Pearl's} original causal hierarchy table, provided to explain our approach to the R-T-B causal taxonomy. \emph{Pearl's} hierarchy is based on the rule: questions at level $i$, $i \in \{ 1, 2, 3\}$ can only be answered if information from level $j\in \{j \geq i\}$ is available. There are three levels: 
	\begin{enumerate}
		\item \emph{Association} (low level); it invokes purely statistical	relationships and require no causal information; formally, it is the conditional probability of event $Y=y$ given event $X=x$, i.e., $P(y|x)=p$;
		\item \emph{Intervention} (intermediate level); it involves not just seeing what is, but changing what we see; formally, it is the probability of event $Y=y$ given that we intervene and set the value $X$ to $x$ and subsequently observe event $Z=z$, i.e., $P(y|\texttt{do}(x)$;
		\item \emph{Counterfactuals} (top level); {a mode of necessitating retrospective reasoning}; formally, the probability of event $Y=y$ would be observed had $X$ been $x$, given that actually observed $X$ to be $x'$ and $Y$ to be $y'$.
	\end{enumerate}
	
	\begin{table}[!ht]
		\caption{The three-layer causal inference hierarchy based on \cite{[Pearl-2019]}.}
		\label{tab:Causal-inference-hierarchy}
		\begin{center}
			\begin{small}
				\begin{tabular}{c|c|c}
					\hline
					\multicolumn{1}{c|}{\begin{parbox}[h]{0.22\linewidth} {\centering
								{Level}}
					\end{parbox}}
					&
					\multicolumn{1}{c|}{\begin{parbox}[h]{0.22\linewidth} { \centering
								{Activity}
					}\end{parbox}}
					&
					\multicolumn{1}{c}{\begin{parbox}[h]{0.3\linewidth} { \centering
								{Typical question}
					}\end{parbox}}
					\\
					\hline
					\begin{parbox}[h]{0.22\linewidth}{
							I.\\ Association\\
							$P(y|x)$
						}
					\end{parbox} 
					&
					\begin{parbox}[h]{0.22\linewidth}{
							\vspace{1mm}
							Seeing
						}
					\end{parbox} 
					&
					\begin{parbox}[h]{0.35\linewidth}{
							\vspace{1mm}
							What is?
							How would seeing $X$ change my belief in $Y$?
							\vspace{1mm}
						}
					\end{parbox} 
					\\
					\hline	
					\begin{parbox}[h]{0.22\linewidth}{
							\vspace{1mm}
							II.\\ Intervention\\
							$P(y|\texttt{do}(x), z)$
						}
					\end{parbox} 
					&
					\begin{parbox}[h]{0.22\linewidth}{
							\vspace{1mm}
							Doing\\
							Intervening
						}
					\end{parbox} 
					&
					\begin{parbox}[h]{0.35\linewidth}{
							\vspace{1mm}
							What if?\\
							What if I do $X$?
						}
					\end{parbox} 
					\\
					\hline	
					\begin{parbox}[h]{0.22\linewidth}{
							\vspace{1mm}
							III.\\ Counterfactuals\\
							$P(y_x |x',y')$
							\vspace{1mm}}
					\end{parbox} 
					&
					\begin{parbox}[h]{0.22\linewidth}{
							\vspace{1mm}
							Imagining,\\
							Retrospection
							\vspace{1mm}}
					\end{parbox} 
					&
					\begin{parbox}[h]{0.35\linewidth}{
							\vspace{1mm}
							Why?
							Was it $X$ that caused $Y$?
							What if I had acted
							differently?
							\vspace{1mm}}
					\end{parbox} 
					\\
					\hline		
				\end{tabular} 
			\end{small}
		\end{center}
	\end{table}
	
	Table \ref{tab:RiskTrustBias} together with Fig. \ref{fig:RiskTrustBias} provide the R-T-B taxonomical view on the intelligent DSS. Three kinds of ensembles are distinguished in the R-T-B space: 
	
	\begin{table*}[!htpb]
		\caption{The R-T-B taxonomy based on \emph{Pearl's} causal inference hierarchy \cite{[Pearl-2019]}}
		\label{tab:RiskTrustBias}
		\begin{center}
			\begin{small}
				\begin{tabular}{c|c}
					\hline
					\multicolumn{1}{c|}{
						{
							\begin{parbox}[h]{0.6\linewidth} {\centering
									\textbf{The R-T-B causal inference}
					}\end{parbox}}}
					& \multicolumn{1}{|c}{{
							\begin{parbox}[h]{0.15\linewidth} {\normalsize \centering
									\textbf{Graph}
					}\end{parbox}}}\\
					\hline
					\hline
					\multicolumn{2}{l}{\normalsize \texttt{I. First order R-T-B}}\\
					\hline
					\hline
					\begin{parbox}[h]{0.65\linewidth}{
							\vspace{1mm}
							\textbf{Risk $R$}: Risk of event $X=x$, \texttt{Risk$=F$(Impact, Probability $X=x$)}\\
							\textbf{Trust} $T$: Trust for event $X=x$\\
							\textbf{Bias} $B$: Bias of event $X=x$
							\vspace{1mm}}
					\end{parbox} 
					&
					\begin{parbox}[h]{0.25\linewidth}{\centering
							\vspace{1mm}
							\includegraphics[scale=0.45]{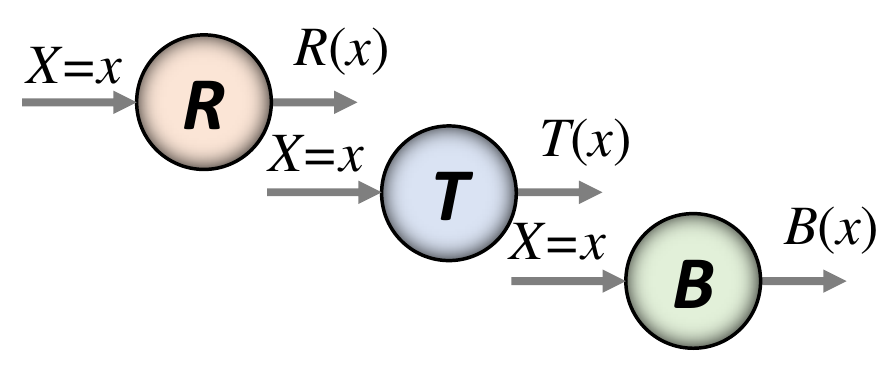}
							\vspace{1mm}}
					\end{parbox}\\
					\hline
					\hline
					\multicolumn{2}{l}{\normalsize \texttt{II. Second order R-T-B}}\\
					\hline
					\hline
					\begin{parbox}[h]{0.6\linewidth}{
							\vspace{1mm}
							\textbf{Risk$\leftarrow$Trust:} 
							\begin{itemize}
								\item [$-$]\emph{Association} $R(x|t)$: Risk of event $X=x$ given that trust $T=t$ is observed;
								\item [$-$]\emph{Intervention} $R(x|\texttt{do}(t,x)$: Risk of event $X=x$ given that we intervene and set the value of trust $T=t$ and subsequently observe event $Z=z$;
								\item [$-$]\emph{Counterfactuals} $R(x_t|t{'},x{'})$: Risk of event $X=x$ would be observed had $X$ been $x$, given that we actually observed $X$ to be $x{'}$ and trust $T$ to be $t{'}$
							\end{itemize}
							\vspace{1mm}}
					\end{parbox} 
					&
					\begin{parbox}[h]{0.2\linewidth}{\centering
							\vspace{1mm}
							\includegraphics[scale=0.45]{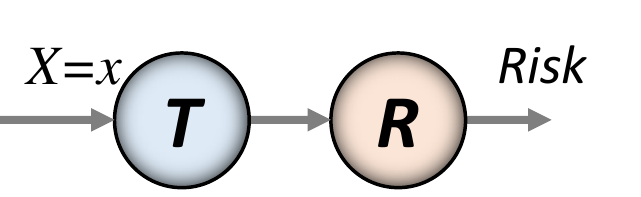}
							\vspace{1mm}}
					\end{parbox}\\
					\hline
					\begin{parbox}[h]{0.6\linewidth}{
							\vspace{1mm}
							\textbf{Risk$\leftarrow$Bias:} 
							
							\emph{Association} $R(x|b)$;
							\emph{Intervention} $R(x|\texttt{do}(b),x)$;
							\emph{Counterfactuals} $R(x_b|b',x{'})$
							
							\vspace{1mm}}
					\end{parbox} 
					&
					\begin{parbox}[h]{0.2\linewidth}{\centering
							\vspace{1mm}
							\includegraphics[scale=0.45]{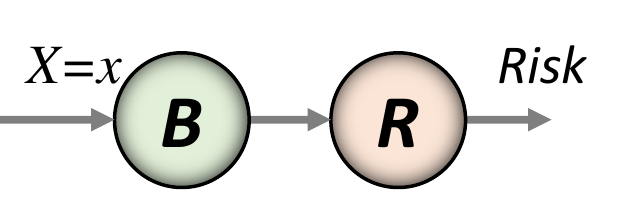}
							\vspace{1mm}}
					\end{parbox}\\
					\hline
					\begin{parbox}[h]{0.6\linewidth}{
							\vspace{1mm}
							\textbf{Trust$\leftarrow$Bias:} 
							
							\emph{Association} $T(x|b)$;
							\emph{Intervention} $T(x|\texttt{do}(b),x)$;
							\emph{Counterfactuals} $T(x_b|b{'},x{'})$
							
							\vspace{1mm}}
					\end{parbox} 
					&
					\begin{parbox}[h]{0.2\linewidth}{\centering
							\vspace{1mm}
							\includegraphics[scale=0.45]{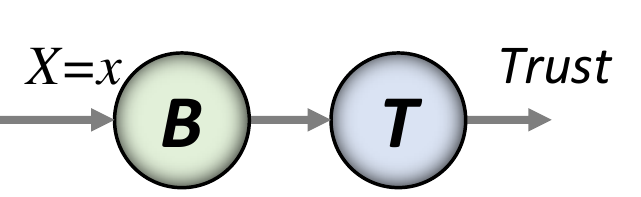}
							\vspace{1mm}}
					\end{parbox}\\
					\hline
					\begin{parbox}[h]{0.6\linewidth}{
							\vspace{1mm}
							\textbf{Trust$\leftarrow$Risk:} 
							
							\emph{Association} $T(x|r)$;
							\emph{Intervention} $T(x|\texttt{do}(r),x)$;
							\emph{Counterfactuals} $T(x_r|r{'},x{'})$
							
							\vspace{1mm}}
					\end{parbox} 
					&
					\begin{parbox}[h]{0.2\linewidth}{\centering
							\vspace{1mm}
							\includegraphics[scale=0.45]{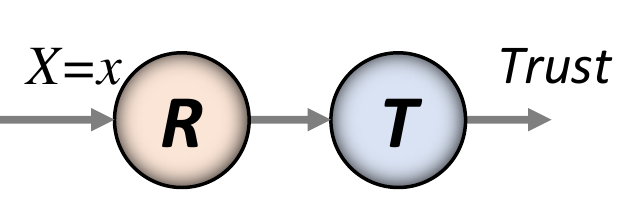}
							\vspace{1mm}}
					\end{parbox}\\
					\hline
					\begin{parbox}[h]{0.6\linewidth}{
							\vspace{1mm}
							\textbf{Bias$\leftarrow$Risk:} 
							
							\emph{Association} $B(x|r)$;
							\emph{Intervention} $B(x|\texttt{do}(r),x)$;
							\emph{Counterfactuals} $B(x_r|r{'},x{'})$
							
							\vspace{1mm}}
					\end{parbox} 
					&
					\begin{parbox}[h]{0.2\linewidth}{\centering
							\vspace{1mm}
							\includegraphics[scale=0.45]{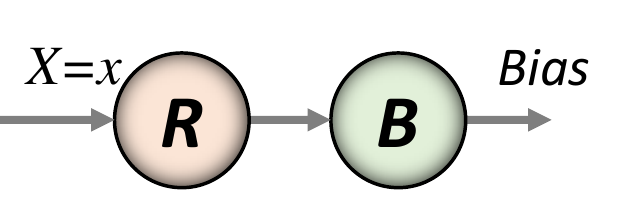}
							\vspace{1mm}}
					\end{parbox}\\
					\hline
					\begin{parbox}[h]{0.6\linewidth}{
							\vspace{1mm}
							\textbf{Bias$\leftarrow$Trust:} 
							
							\emph{Association} $B(x|t)$;
							\emph{Intervention} $B(x|\texttt{do}(t),x)$;
							\emph{Counterfactuals} $B(x_t|t{'},x{'})$
							
							\vspace{1mm}}
					\end{parbox} 
					&
					\begin{parbox}[h]{0.2\linewidth}{\centering
							\vspace{1mm}
							\includegraphics[scale=0.45]{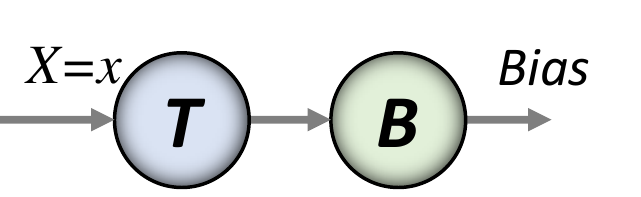}
							\vspace{1mm}}
					\end{parbox}\\
					\hline
					\hline
					\multicolumn{2}{l}{\normalsize \texttt{III. Third order R-T-B}}\\
					\hline
					\hline
					\begin{parbox}[h]{0.6\linewidth}{
							\vspace{1mm}
							\textbf{Bias$\rightarrow$Risk$\leftarrow$Trust:} 
							
							\emph{Association} $R(x|b,t)$;
							\emph{Intervention} $R(x|\texttt{do}(b,t),x)$;\\
							\emph{Counterfactuals} $R(x_{b,t}|b{'},t{'},x{'})$
							\vspace{1mm}}
					\end{parbox} 
					&
					\begin{parbox}[h]{0.2\linewidth}{\centering
							\vspace{1mm}
							\includegraphics[scale=0.5]{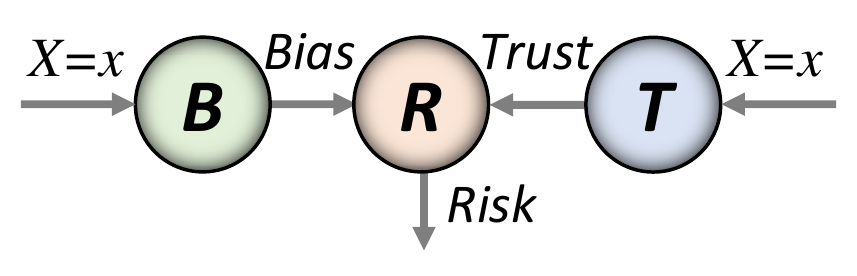}
							\vspace{-2mm}}
					\end{parbox}\\
					\hline
					\begin{parbox}[h]{0.6\linewidth}{
							\vspace{1mm}
							\textbf{Bias$\rightarrow$Trust$\leftarrow$Risk:}
							
							\emph{Association} $T(x|b,r)$;
							\emph{Intervention} $T(x|\texttt{do}(b,r),x)$;\\
							\emph{Counterfactuals} $T(x_{b,r}|b{'},r{'},x{'})$
							\vspace{1mm}}
					\end{parbox} 
					&
					\begin{parbox}[h]{0.2\linewidth}{\centering
							\vspace{1mm}
							\includegraphics[scale=0.5]{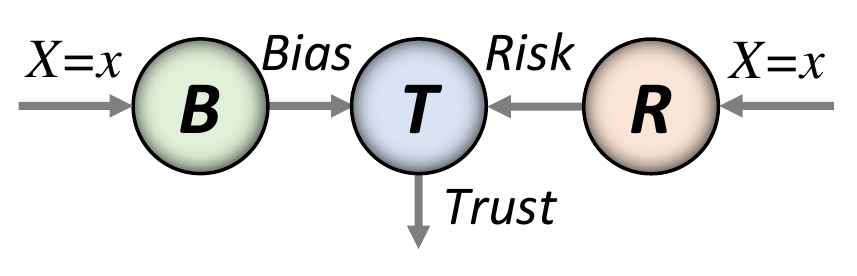}
							\vspace{-2mm}}
					\end{parbox}\\
					\hline
					\begin{parbox}[h]{0.6\linewidth}{
							\vspace{1mm}
							\textbf{Trust$\rightarrow$Bias$\leftarrow$Risk:}\\
							\emph{Association} $B(x|t,r)$;
							\emph{Intervention} $B(x|\texttt{do}(t,r),x)$;\\
							\emph{Counterfactuals} $B(x_{t,r}|t{'},r{'},x{'})$
							\vspace{1mm}}
					\end{parbox} 
					&
					\begin{parbox}[h]{0.2\linewidth}{\centering
							\vspace{1mm}
							\includegraphics[scale=0.5]{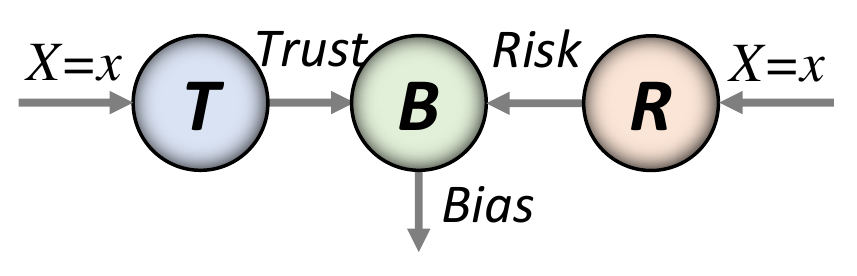}
							\vspace{-2mm}}
					\end{parbox}\\
					\hline		
				\end{tabular} 
			\end{small}
		\end{center}
	\end{table*}

	\emph{1st order:} The R-T-B ensemble represents the simplest (idealized) scenarios of performance regulators, e.g., risk assessment, $R$, ignoring trust and bias factors.
	
	\emph{2nd order:} The R-T-B ensemble reflects the simplest causal relationships, that is, knowledge about the 1st order ensemble, e.g., risk caused by trust $R(T)$.
	
	\emph{3rd order:} The R-T-B ensemble contains knowledge about the 2nd order ensemble, e.g., risk of trust in the presence of bias, $R(T|B)$. For example, in \cite{[Zanin-2013]}, the risk of crime has been studied over social trust in the presence of expected biased parameter estimates. Note that the 3rd order R-T-B include $T(B|R), T(R|B), B(T|R),$ and $B(R|T)$.
	
	In \cite{[Rass-2013]}, trust is derived from experience while assuming multiple ``sources'' of trust in a system. The fusion of trusted sources represented by Beta-distributions is performed using \emph{copula} technique resulting in a joint probability distribution of the overall trust. Paper \cite{[Rass-2013]} then proposes a method for risk forecasting based on a trust model using a mix of Beta-distributions. It takes positive or negative security incident indications as an input and compiles a numerical value within [0,1] that models the probability of system failure, conditional on the recorded experience. The trust model is updated in a Bayesian fashion, making the trust measure a conditional expectation of a security indicator, based on prior experience.
	
	Summarizing, Table \ref{tab:RiskTrustBias} outlines a causal R-T-B relationships. These can be inferred using various types of causal graph models, a brief guide to which is provided in the next Section.

	\section{Systematic view of advanced causal networks}
	\label{sec:Syst-view-causal-nets}
	
	The crucial requirement for the R-T-B formalization is the ability to reason about the R-T-B state as well as the R-T-B prediction. Specifically, the following conditional dependencies can be derived from the R-T-B causal landscape in Fig. \ref{fig:RiskTrustBias}:
	
	\begin{itemize}
		\item [$(a)$] Given the bias $B$, assess/predict the risk of trust 
		${R(T|B)}$ and trust of risk ${T(R|B)}$;
		\item [$(b)$] Given the trust $T$, assess/predict the risk of bias
		${R(B|T)}$ and bias of risk ${B(R|T)}$; 
		\item [$(c)$] Given the risk $R$, assess/predict the trust of bias
		${T(B|R)}$ and bias of trust ${B(T|R)}$.
	\end{itemize}
	
	A causal network is a directed acyclic graph where each node denotes a unique random variable. A directed edge from node \(n_1\) to node \(n_2\) indicates that the value attained by \(n_1\) has a direct causal influence on the value attained by \(n_2\). Uncertainty inference requires a specific type of data structures referred to as \emph{Conditional Uncertainty Tables} (CUTs). A CUT is assigned to each node in the causal network. Given a node \(n\), the CUT assigned to \(n\) is a table that is indexed by all possible value assignments according to the parent nodes of \(n\). Each entry of the table is a conditional ``uncertainty model'' that varies according to the choice of the uncertainty metric.
	
	Analysis of a causal network is out of the scope of this paper. However, we introduce in this paper the \emph{systematic criteria} for choosing the appropriate computational tools. In addition, some details are clarified in our experimental study. 
	
	The following types of causal networks are deployed in contemporary machine reasoning based on the CUT criterion:
	
	\begin{small}
		\begin{center}
			\begin{parbox}[h]{0.95\linewidth} {
					\vspace{-2mm}
					\begin{center}
						\begin{eqnarray}\label{Review}
						\begin{parbox}[h]{0.16\linewidth} {\centering
							\textbf{\texttt{Causal network}} }
						\end{parbox}
						\equiv\left\{\hspace{-0.2cm}
						\begin{tabular}{lll}
						\texttt{\small Bayesian} & {\small CUT$\equiv$CPT} & 
						\cite{[Pearl-2000]}; \\
						\texttt{\small Imprecise} & {\small CUT$\equiv$CImT}& \cite{[Coolen-2011]}; \\
						\texttt{\small Interval} & {\small CUT$\equiv$CInT}&\cite{[DeCampos1994]};\\
						\texttt{\small Credal} & {\small CUT$\equiv$CCT}&\cite{[Cozman-2000]}; \\
						\texttt{\small DS } & {\small CUT$\equiv$CDST}& \cite{[Simona-2008]};\\
						\texttt{\small Fuzzy} & {\small CUT$\equiv$CFT}&\cite{[Baldwin-2003]}; \\
						\texttt{\small Subjective} & {\small CUT$\equiv$CST}&\cite{[Ivanovska-2015]}; \\
						\end{tabular} 
						\hspace{-0.2cm}\right \}
						\end{eqnarray}
				\end{center}} 
			\end{parbox}
		\end{center}
	\end{small}
	
	In the list (\ref{Review}), the following abbreviations are used: 
	
	\begin{tabular}{lll}
		CPT&--& Conditional Probability Table;\\
		CImT &--& Conditional Imprecise Table;\\
		CInT &--& Conditional Interval Table;\\
		CCT &--& Conditional Credal Table;\\
		CDST &--& Conditional Dempster-Shafer (DS) Table;\\
		CST &--& Conditional Subjective Table.
	\end{tabular} 
	
	The distinguishing feature of these CUTs is that the uncertainty is interpreted in different ways. For example, uncertainty in risk assessment can be ``filled'' by weighted compositions of costs of losses, or by a set of alternative decisions.	The type of a causal network shall be chosen given the DSS model and a specific scenario. The choice depends on the CUT as a carrier of \emph{primary} knowledge and as appropriate to the scenario:
	\begin{itemize}
		\item []\hspace{-8mm} \emph{Bayesian network} is defined as a causal network with the CUT being CPT using point probability measures \cite{[Pearl-2019]}. The key limiting factor is the assumption that modeled events are independent.
		\item []\hspace{-8mm} \emph{Imprecise causal network} is defined by using the CUT type such as the CImT, using lower and upper probabilities $\overline{p}(A)$ and $\underline{p}(A)$, respectively \cite{[Coolen-2011]}.
		\item []\hspace{-8mm} \emph{Interval causal network} is defined by the specification of the CUT as the CInT and probability interval using a ``radius of uncertainty'' for each point probability \cite{[DeCampos1994]}.
		\item []\hspace{-8mm} \emph{Credal causal network} is defined by specifying the CUT as the CCT using closed intervals of the possible range of probability values \cite{[Cozman-2000]}; this model can be viewed as a set of Bayesian networks that share the same graphical structure but are associated with different conditional probability parameters.
		\item []\hspace{-8mm} \emph{Dempster-Shafer (DS) causal network} is defined by using the CDST that utilizes the formalization of \emph{imprecise} probabilities for evaluating the quality of results, producing optimistic and pessimistic estimations of vulnerability via \emph{plausibility} and \emph{belief} measures	\cite{[Simona-2008]}.
		\item []\hspace{-8mm} \emph{Fuzzy causal network} is defined by specifying the CFT using fuzzy measures \cite{[Baldwin-2003]}. The CFTs are similar to CInT, but the lower and upper bounds may be ``soft''.
		\item []\hspace{-8mm} \emph{Subjective causal network} is defined by specifying the CUT as the CST using subjective opinions, a belief-and-uncertainty representation of an unknown probability distribution of a random variable \cite{[Ivanovska-2015]}.
	\end{itemize}
	
	The choice of a specific causal network model is heavily dependent on the data that is available for creating the CUTs, as well as the information that is expected to be given by the posterior uncertainty model. For instance, if statistical data is in abundance, {probability theory} will be the most suitable choice of uncertainty model and will provide the most informative results. If statistical data is lacking for certain variables, probability intervals can account for uncertainty in those probabilities for which there is insufficient data. If statistical data is almost completely lacking, DS theory may be appropriate and the expert can provide the DS weights to populate the CDSTs. 
	
	Note that specific biases can be observed in reasoning using causal networks, such as \emph{endogeneity bias}. Endogeneity occurs when an omitted variable or a variable's value confounds the relationship between cause and effect, thereby introducing bias into the estimate of the causal effect and reasoning mechanism. In statistical terms, the endogeneity of a given variable manifests itself as an association between the variable and the error term. For example, in \cite{[Zanin-2013]}, the societal R-T-B assessments have been considered with respect to the endogeneity bias.
	
	There were several attempts to provide researchers with the \emph{``Guidelines''} for choosing the best causal network platform based on the CUT. A recent review \cite{[Rohmer-2020]} covers most of the network types in list (\ref{Review}). Comparison of causal computational platforms for modeling various systems is a useful strategy, such as Dempster-Shafer vs. credal networks \cite{[Misuri-2018]}, Bayesian vs. interval vs. Dempster-Shafer vs. fuzzy networks \cite{[Yanushkevich-2018],[Yanush-2019-a]}. 	
	
	R-T-B reasoning is the ability to form an intelligent judgment using the R-T-B data. It is a judgment under uncertainty based on a causal network. For example, in \cite{[Hunter-2017]}, the notion of trust is closely connected to the notion of belief change. Trust is defined in terms of a trust partition over a set of belief states, and the belief is updated based on the trust-sensitive belief revision operators.

	\section{Standardization of the R-T-B measures using Admiralty Code}\label{sec:Admiralty-Code}
	
	R-T-B and their causal relationships manifest themselves in intelligent DSS in different ways, such as the reliability of information (data) sources and the credibility of the information (data):
	
	\begin{footnotesize}
		\begin{center}
			\begin{parbox}[h]{0.95\linewidth} {
					\vspace{-2mm}
					\begin{center}
						$\underbrace{\texttt{Source~Reliability}}_{\textit{Quality}}~{\Leftrightarrow}
						\left\{\hspace{-0.2cm}
						\begin{array}{c}
						\texttt{Risk} \\
						\texttt{Trust} \\
						\texttt{Bias} \\
						\end{array}\hspace{-0.2cm}
						\right\}{\Leftrightarrow}~
						\underbrace{\texttt{Data~Credibility}}_{\textit{Reputation}}$\\
					\end{center}
			} \end{parbox}
		\end{center}
	\end{footnotesize}
	
	The relationship can be represented as follows:
	\begin{itemize}
		\item [$(a)$]\emph{Source reliability} as the quality of being reliable, or trustworthy, is related to 1) risk as a function of potential adverse impact and the likelihood of occurrence, 2) trust as the confidence in quality, as well as their causal relationships, and 3) bias as systematic over- or under-assessment of the parameter of interest.
		\item [$(b)$]\emph{Information (data) credibility} as the reputation impacting one's ability to be believed. 
	\end{itemize}
	
	In \cite{[Yanushkevich-2018]}, available resources and information for traveler profiling are rated accordingly to the Admiralty Code (Fig. \ref{fig:Admiralty-Code}). NATO Standardization Agreements such as STANAG 2022 and STANAG 2511 \cite{[Admiralty_Code_2012],[U.S._Army_Field_Manual],[Blasch-2013]} use the Admiralty Code to resolve conflicting scenarios in human-human, human-machine, and machine-machine interactions. In \cite{[Blasch-2013]}, information trust is defined based on well-formalized reliability and credibility attributes. 
	
	The reliability of the decision support provided by the DSS can be increased by using more reliable sources and credible information or can be diminished due to lowered reliability of the source and/or credibility of the information. In this context, trust can be expressed in terms of the reliability of data sources and/or the credibility of prior information. For example, scenario $F6$ is composed of the source reliability \texttt{F$\equiv$\ <Cannot be judged>} and information credibility \texttt{6$\equiv$\ <Cannot be judged>}. 
	
	There are various ways to use the Admiralty Code standard. For example, notion ``credibility'' is equivalent to ``trustworthiness'' over ``expertise'' where ``trustworthiness'' represents a fused reliability of source and credibility of data.
	
	Fig. \ref{fig:Admiralty-Code} explains the decision support mechanism for assessing different scenarios in terms of \emph{system states}. For example, given the states $\{S_{1}, S_{2},\ldots , S_{8}\}$ of the epidemiological surveillance, the DSS analyzes the states according to the Admiralty Code resulting in the following decision-making landscape: 
	
	\begin{itemize}
		\item [$-$]States $S_{1}$ and $S_{8}$; $S_{2}$ and $S_{7}$ can be used for decision-making;
		\item [$-$]Decision-making based on states $S_{3}$, $S_{4}$, and $S_{5}$; $S_{6}$ is very risky.
	\end{itemize}
	
	
	\begin{figure}[!ht]
		\begin{center}
			\includegraphics[width=0.5\textwidth]{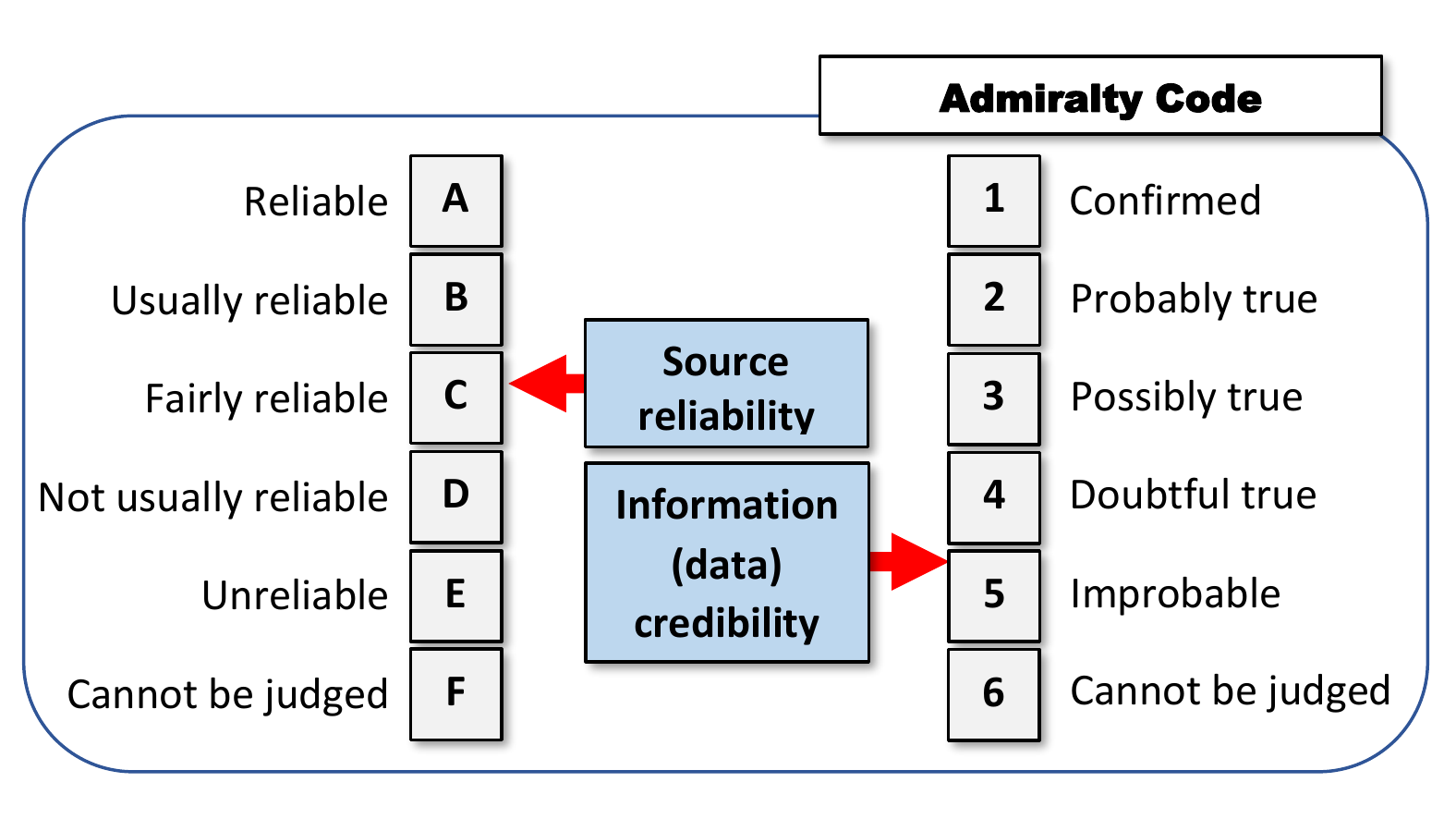}
		\end{center}
		\caption{Manifestation of the R-T-B via assessments of reliability of source and credibility of information using the Admiralty Code.}
		\label{fig:AdmiraltyCode}
	\end{figure}
	
	\begin{figure}[!ht]
		\begin{center}
			\begin{footnotesize}
				\begin{eqnarray*}
					\bordermatrix{
						& \mbox{$1$} & \mbox{$2$} & \mbox{$3$} & \mbox{$4$}& \mbox{$5$}& \mbox{$6$}\cr \noalign{
							\vskip 2pt}
						\mbox{~$A$}& &\fbox{$S_{2,7}$} && & &\cr \noalign{
							\vskip 2pt}
						\mbox{~$B$}& & & & &&\cr \noalign{
							\vskip 2pt}
						\mbox{~$C$}&\fbox{$S_{1,8}$}& & &&\fbox{$S_{3,4,5}$} &\cr \noalign{
							\vskip 2pt}
						\mbox{~$D$}& & & & & &\cr \noalign{
							\vskip 2pt}
						\mbox{~$E$}& &\fbox{$S_{6}$}& & & &\cr \noalign{\vskip 2pt}
						\mbox{~$F$}& & & & & &\cr \noalign{
							\vskip 2pt}
					}
				\end{eqnarray*}
			\end{footnotesize}
		\end{center}
		\caption{Example of the R-T-B reasoning. Given a set of a system states $\textbf{S}=S_i,~i=1,2, \ldots , 8.$ The primary task of the DSS is the R-T-B reasoning about these states using available resources such as the Admiralty code. The result is a set system states provided to a human analyst/expert to support their decision-making. }
		\label{fig:Admiralty-Code}
	\end{figure}
	
	For security checkpoints, the source reliability and information credibility are represented by the probabilistic variables such as false ID, multiple ID of the same person, and features of intentional data alteration in the chip (e.g., biometric traits and text data), as well as a false life-cycle history~\cite{[Andreou-2017],[Yang-2013]}.

	\section{Demonstrative experiments}
	\label{sec:Exp}
	
	The goal of this section is to demonstrate how the R-T-B concept works in real-world large scale tasks, and through this method, empirically prove that the R-T-B triad is a system performance regulator. For this, a typical biometric-enabled complex dynamic system was chosen. Two research questions was prioritized for investigation in the R-T-B dimensions: 
	\begin{itemize}
		\item [$-$] Impact of synthetic data on system performance; this problem is motivated, in particular, by research \cite{[Gavrilova-2011],[Bellovin-2019]}; 
		\item [$-$] Impact of demographic factors on system performance; this problem is critical in facial recognition \cite{[Das-2019],[Grother-2019]}.
	\end{itemize}
	
	\subsection{The R-T-B of synthetic data}	
	
	Experimental study of synthetic data impact on performance of a cognitive checkpoint is reported in \cite{[Yanush-2020-b]}. Below, we briefly introduce the quintessence of this report and provide new projections. 	
	\subsubsection{Problem}	 
	Synthetic data often replaces authentic data or is used together with the latter. They are an essential part of modeling and training various components of a checkpoint. Synthetic biometric traits are a class of algorithmically 	generated biometric, non-biometric, and cognitive behavior authentication credentials (e.g., face and facial expressions, fingerprints, voice, gait, user name, password, ) used as a source for constructing a human profile for identity management \cite{[Active-Authen-DARPA-2016]}.
	
	\subsubsection{Multi-state screening model}	
	
\begin{figure}[!ht]
		\begin{center}
			\includegraphics[width=0.5\textwidth]{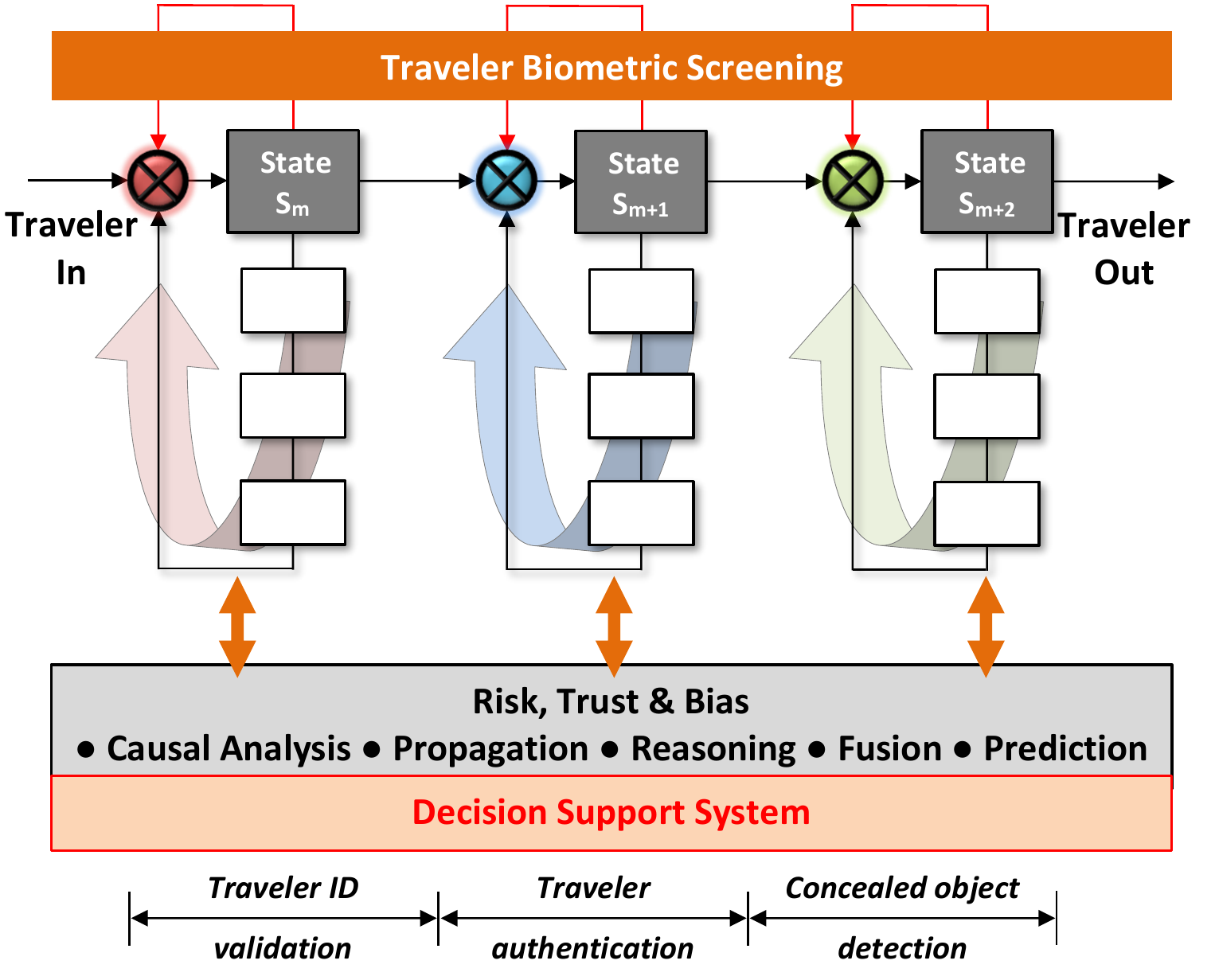}
			\caption{The taxonomical view of the multi-state intelligent identity management process. The R-T-B of synthetic data is assessed considering their causal relationships with authentic data. Each state is represented by a perception-action cycle of sub-states. }
			\label{fig:Checkpoint}
		\end{center}
	\end{figure}

	We consider a multi-state screening in the dynamic cognitive system that (Fig. \ref{fig:Checkpoint}): 
	\begin{enumerate}
		\item Monitors the traveler data throughout the process of e-ID checking, face recognition, and continuously assess the R-T-B using various sources such as behavioral biometrics, watchlist, and CI decision assistant results;
		\item Updates its states based on the intelligence gathered via human-machine interactions (CI decision assistant), the results of the biometric traits recognition based on machine learning, the results of the concealed object detection (by adjusting radar illumination), and others.
	\end{enumerate}
	
	In Fig. \ref{fig:Checkpoint}, the traveler's identity management process is implemented in three states, $S_1$ (ID validation), $S_2$ (Traveler authentication), and $S_3$ (Concealed object detection). Each state $S_i$ and sub-state is a part of the `Layered Security Strategy', a contemporary security doctrine \cite{[TSA-2013]}. Each state $S_i$ and sub-state generates the R-T-B assessments for further processing. Inference using operations such as propagation, causal analysis, and reasoning can also be applied to the R-T-B assessments.
	
	Because R-T-B are measured as probability events, they can be combined using \emph{propagation} and \emph{fusion} techniques. Synthetic data is required at various CI {operations} and processes. For example, the sub-state $S_{m}^{(1)}$ of state $S_{m}$ is defined under \emph{learnt} ID source reliability using authentic data from previous experience, while the potential attack data can be synthesized. This enables assessment of the R-T-B of such rare events (attacks). 
	
	\subsubsection{Formalization accordingly the Admiralty Code}
	Consider a typical real-world scenario of the ID management process: Given an e-ID, {assess} the ID information credibility. At the descriptive level of the Admiralty Code, this scenario is represented as (Section \ref{sec:Admiralty-Code}):
	\begin{center}
		\texttt{\small<Credibility>} $\equiv$ 
		\texttt{\small<Trustworthiness>}
		+\texttt{\small<Expertise>}
	\end{center}
	where \texttt{\small<Trustworthiness>} manifests itself as 
	\begin{eqnarray*}
		{\texttt{\small Trustworthiness}} 
		\equiv \left\{
		\begin{tabular}{l}
			\texttt{\small Source~Reliability} \\
			$+$ \\
			\texttt{\small Data~Credibility} \\
		\end{tabular} 
		\right .
	\end{eqnarray*}
	
	\subsubsection{Causal network}
	
\begin{figure}[!ht]
		\begin{center}
			\includegraphics[scale=0.55]{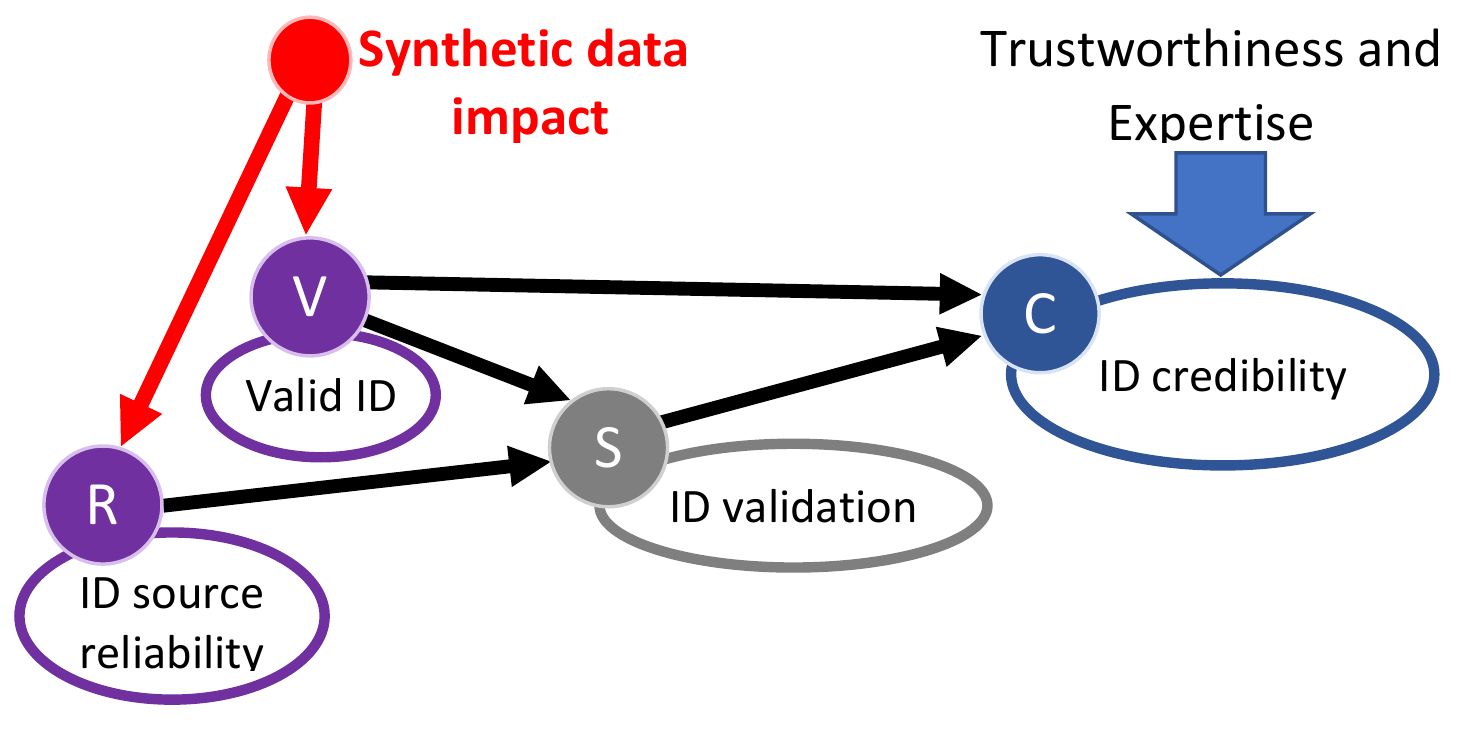}
		\end{center}
		\caption{Assessment of the ID credibility (trustworthiness and expertise) using an IV-echelon (state) identity management scenario and its implementation. Synthetic data impact is incorporated using the CPTs of the nodes $R$ and $V$. }\label{fig:BayesianNet}
	\end{figure}

	Assessment of the ID information credibility is represented in Fig. \ref{fig:BayesianNet} in the form of a Bayesian network and the corresponding CPTs are as follows:
	
	\begin{itemize}
		\item []\hspace{-8mm} Node \emph{`ID source reliability'} denotes the three reliability levels of the e-passport/ID authentication, which depends on many risk factors such as the country of issue, the number of defence levels in the document, the life cycle history, the type of chip, the type of biometric modality, the type of encryption, and the type of RFID mechanism.
		\item []\hspace{-8mm} Node \emph{`Valid ID'}, or \emph{`Trusted ID'} denotes whether the e-passport ID should pass the validation procedure using factors such as watermarks, holograms, ultraviolet threads, micro text, and optical variable ink.		
		\item []\hspace{-8mm} Node \emph{`ID validation'} denotes the outcome of the authentication process of the e-passport. 
		\item []\hspace{-8mm} Node \emph{`ID credibility'} describes the three credibility level of the
		outcome of the validation process. If the credibility of the validation process is known as a \emph{priori}, it can be used to compute the posterior beliefs related to the validity of the individual document (node \(V\)).
	\end{itemize}
	
	\subsubsection{Scenario and reasoning example}
	
	Consider the following particular scenario:	{IF the reliability of the} {ID source } {is } {known } {to be } {`low' and the } {resulting} {credibility} is `high', {THEN} what is the posterior probability that the ID is valid? This scenario models a situation of conflict where an unreliable source produces a credible outcome. It is very likely that the ID was valid. That is, the trustworthiness of the statement `the ID was valid' is coherent with the expert knowledge (incorporated in the algorithms) \cite{[Yanush-2020-b]}.

	\subsubsection{Synthetic data risks}
	Let us assume that to train algorithms for validating ID (node~$V$) and identifying ID source reliability (node $R$), synthetic data was used to represent rare events, such as false ID, multiple ID of the same person, and features of intentional data {alteration} in the chip (e.g., biometric traits and text data) as well as a false life cycle history. Probabilities of these threats are represented by the CPTs for nodes $V$ and $R$. There is always a risk that the validation algorithm makes a mistake should the real rare event occur. For example, features of the forged e-ID are not detected, or these features can be mistakenly detected in a valid ID. The goal is to assess these risks caused by the usage of synthetic data.
	
	It is well understood that the frequency of object occurrence in the identity management process follows a long-tailed distribution. For example, people with true IDs and expired IDs are much more common than people with false IDs and multiple IDs. This problem relates the novelty detection, also known as \emph{anomaly detection}, or one-class classification, the task of recognizing that the test data differ in some respect from the data that are available during training \cite{[Pimentel-2014]}. The tailed probability distributions have been used, for example, in the study of cyber-risks such as ID theft \cite{[Maillart-2010]}.
	
	\subsection{Bias ensemble in facial recognition}
	\label{sec:experiment_II}
	
	There are three phases of bias analysis in the DSS: 1) Bias identification, e.g., what kind of biases are manifested in a system? 2) Bias assessment, e.g., a unified metric for different kinds of biases, such as the risk of bias; and 3) Bias operation, e.g., the fusion of risks of biases.
	
	\subsubsection{Problem}
	\label{sec:Taxonomical-projection}
	
	The traveler's identity management process is implemented as a process that goes through multiple states \cite{[Yanush-2019],[Yanush-2019-a]}. Each state is characterized by a specific bias such as the bias in face recognition. Statistics of these biases are being used for machine learning. These biases are mostly represented by the tails of the probability distributions. A unified metric of bias that we consider in this study is the risk of bias and the related trust in the technology that is biased.
	
	Our experiment addresses multiple biases in a cognitive security checkpoint such as ``ID Reliability Bias'', ``ID Validation Bias'', and ``Trustworthiness Bias''. Among various candidates of biases, we consider specifically ``Face Recognition Bias''. Few results on demographic bias in facial recognition have been recently reported, in particular, in \cite{[Das-2019],[Grother-2019],[Merler-2019]}. The experimental details are described in \cite{[Lai-2020]}, which aims to highlight the practical details of assessing an ensemble of biases. 
	
	\subsubsection{Causal model}
	
\begin{figure}[!ht]
		\begin{center}
			\includegraphics[width=0.45\textwidth,interpolate]{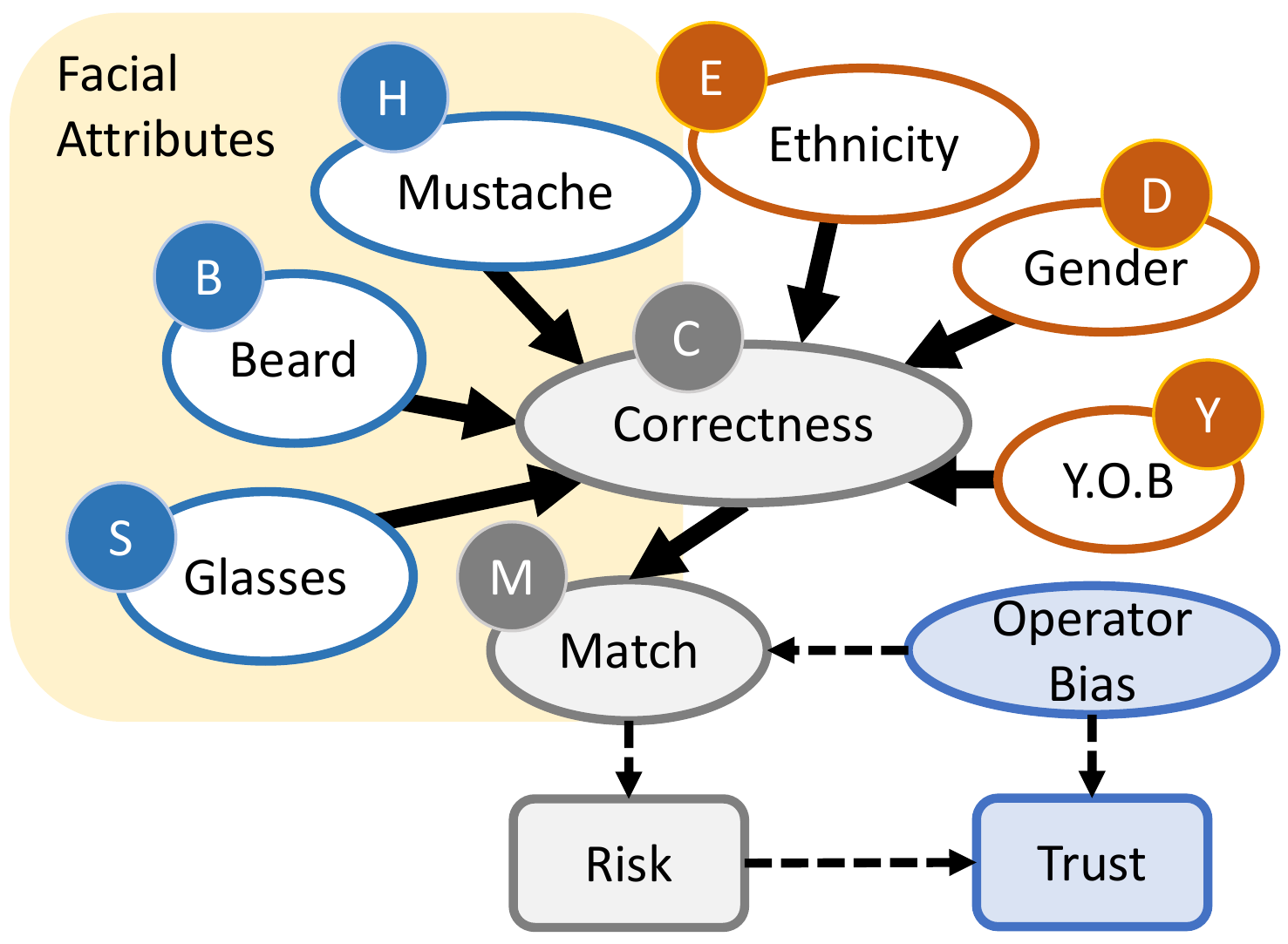}
			\caption{A simplified causal network of biases in facial recognition. Risk is derived based on the results of the ``Match'', and Trust is affected by the Operator's Bias. }
			\label{fig:Biases}
		\end{center}
	\end{figure}

	The causal network shown in Fig. \ref{fig:Biases} describes how the quality of facial recognition can be compromised by the various facial attributes that are ``biased'' based on the year-of-birth (YOB) $Y$, gender $D$, ethnicity $E$, mustache $H$, beard $B$, and glasses $S$. The parent nodes to the ``Correctness'' node represents the bias attributed to face recognition. The ``Correctness'' node presents the probability of the neural network in predicting a positive (genuine subject) or negative (imposter) identity, whereas the ``Match'' node determines whether the positive or negative prediction matches the ground truth label.
	
	\subsubsection{Formalization}
	{Risk of error in the decision due to bias is estimated as \texttt{Risk$=F$(Impact, Probability)} which relates to the error rates of the system, specifically the false non-match rate (FNMR) and false match rate (FMR). In addition, the risk value is associated with the probability of a random user being genuine given a particular bias \texttt{P(Genuine|Bias)}. At a high level of abstraction (e.g., ignoring metric and dependencies), the risk given a particular bias $\texttt{Risk}_\text{\slshape Bias}$ is defined as follows:}
	
	\vspace{-2mm}
	\begin{small}
		\begin{eqnarray} \label{eq:riskb}
		\texttt{Risk}_\text{\slshape Bias}\hspace{-2mm}&=&\hspace{-2mm} \underbrace{\texttt{Impact}_{\text{\slshape FMR}}}_\text{Cost of a FMR} \times \texttt{FMR}_{\text{\slshape Bias}} \times (1-\texttt{P(Genuine|Bias)})\nonumber\\
		&+&\underbrace{\texttt{Impact}_{\text{\slshape FNMR}}}_\text{Cost of a FNMR} \times \texttt{FNMR}_{\text{\slshape Bias}} \times \texttt{P(Genuine|Bias)} \nonumber
		\end{eqnarray}
	\end{small}
	
	\vspace{-3mm}
	For example, given the scenario of a security checkpoint, the FMR is related to a wrongly granted access, while the FNMR contributes to travelers' inconvenience. The impact of the FMR is a breach of security which, given this scenario, should have a high impact. The impact of the FNMR is a negative user experience, which is of a lower impact. Based on this scenario, we assign a 10:1 impact ratio and a 90\% genuine user probability, that is $\texttt{Impact}_{FMR}=10$, $\texttt{Impact}_{FNMR}=1$, $\texttt{P(Genuine|Bias)}=0.9$. Given the YOB attribute, the risk for individuals born in the 1930s is computed as follows: $10 \times 0.0208 \times 0.1 + 1 \times 0.0012 \times 0.9 = \fbox{0.02188}$.
	
	The ensemble risk bias in identifying (matching) a particular individual $\texttt{Risk}_{\text{\slshape Bias}}\texttt{(Ensemble)}$ is assessed as the sum of the risk biases according to his/her attributes:
	
	\vspace{-3mm}
	\begin{eqnarray} \label{eq:risk}
	\texttt{Risk}_{\text{\slshape Bias}}\text{(\texttt{Ensemble})}&=& \sum_{\text{\slshape Bias}=1}^{N} \texttt{Risk}_\text{\slshape Bias}
	\end{eqnarray}
	where $\text{\slshape Bias}$ represents one of the attributes, $Y,D,E,H,B,$ and $S$ in Fig. \ref{fig:Biases}, $N=6$.

	\subsubsection{Experimental results}
	
	For this experiment, we demonstrate biases using the FERET face database that contains a total of 14,126 images of 1199 subjects \cite{phillips1998feret}. The typical performance measure for face recognition includes accuracy, FNMR, and FMR. The features used for face identification is extracted using a pre-trained Resnet50 convolutional neural network. 
	
	The numerical results are presented in detail in \cite{[Lai-2020]}.	The identified biases include gender, year-of-birth, ethnicity, and facial attributes (glasses, beard, and mustache). The causal relationship between the biases and face recognition accuracy is represented by the causal network shown in Fig. \ref{fig:Biases}. A significant bias in face recognition accuracy was observed with respect to the year-of-birth: \textbf{the accuracy decreased by $\mathbf{17.65\%}$ between those born in the 1920s and the 1980s}. 
	
	%
	%
	%
	
	\section{Forecast of emerging applications}
	\label{sec:Special-Section}
\begin{figure}[!ht]
		\begin{center}
			\includegraphics[scale=0.52]{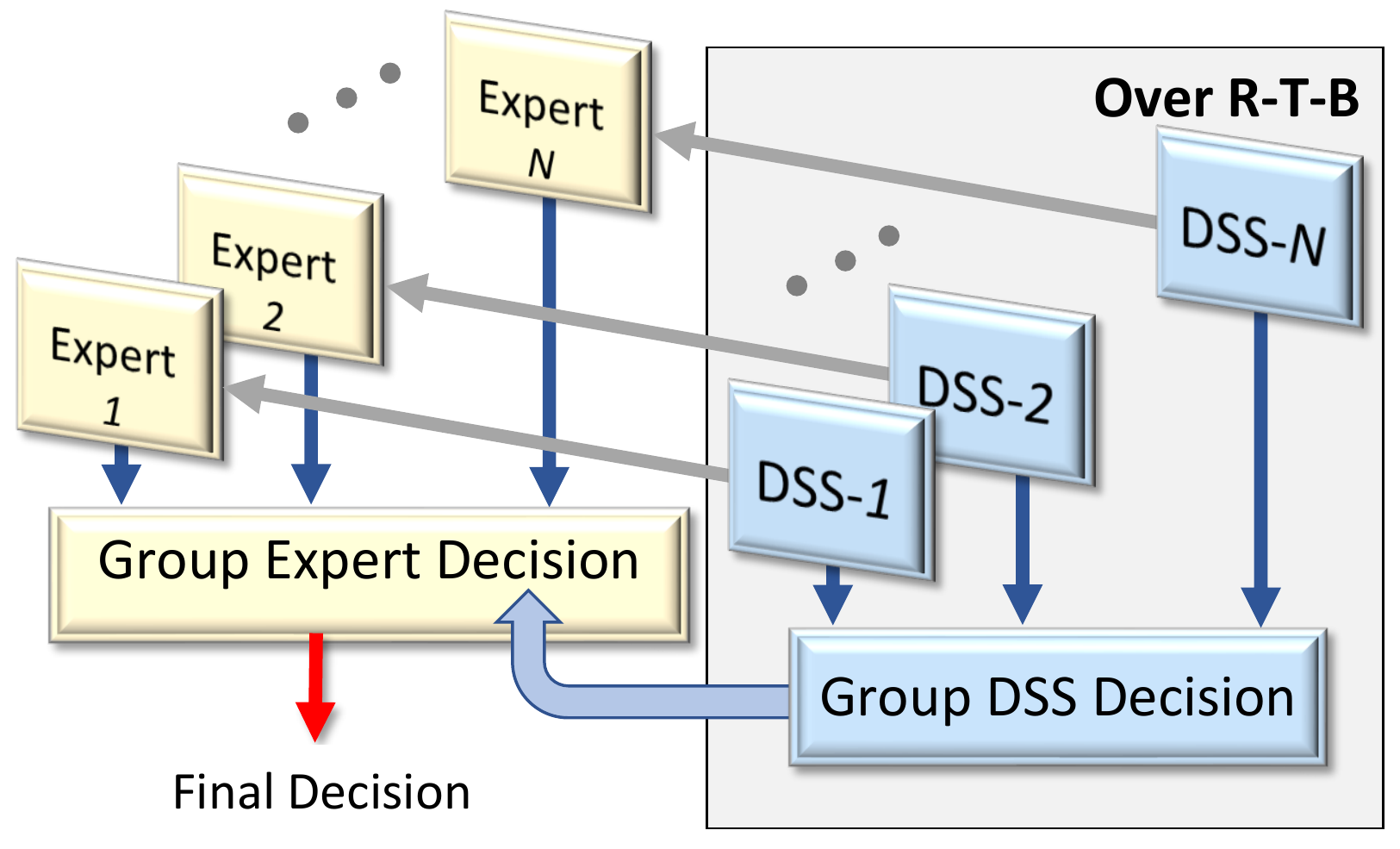}
		\end{center}
		\caption{The framework of the DSS emerging applications over the R-T-B measures: human-machine using the DSS. The Final decision is the consensus of each expert's decision supported by the consensus of the DSS.}
		\label{fig:Group-DSS}
	\end{figure}

	Results of our work are common among processes that can be modeled based on the principles of complex dynamic systems, e.g., learning, teaching, observation, conflict resolving, proactive computations, and countermeasures in real world and cyberspace. Below we introduce several emerging DSS over R-T-B applications:
	
	\begin{itemize}
		\item [$-$] Epidemiological surveillance 
		\cite{[Alamo-2020],[Fenton-2020],[McLachlan-2020],[Neil-2020]};
		\item [$-$] DSS for autonomy systems \cite{[Danks_2017],[deVisser-2018]};
		\item [$-$] Combat DSS \cite{[Brown-2020],[Schmidt-2019],[Pandey-2016]};
		\item [$-$] Ambient DSS \cite{[Li-2015]};
		\item [$-$]	E-coaching	\cite{[Kamphorst-2017],[Ochoa-2018]}.	
	\end{itemize}
	These and other potential applications are based on the concept of group decision-making \cite{[Bedford-2013],[Kamisa-2018]}. Given an evidence and $N$ experts, each expert is supported by the DSS to make a group decision (illustrated in Fig. \ref{fig:Group-DSS}).
	
	\subsubsection{Epidemiological surveillance}
	Epidemiological surveillance experts (healthcare, first response, transportation, education, business and communication, media, security, police, etc.) need a near real-time, accurate picture of the extent and patterns of disease transmission at the community-level. Each expert in a specific field needs an {intelligent DSS} in order to better understand current and evolving healthcare demands, in order to be able to make {low-risk}, time-sensitive rapid decisions over various kinds of {biases} related, for example, on how to allocate limited and/or secure additional resources and how to relax mitigation efforts \cite{[Manheim-2016]}. Part of these tasks uses biometrics, e.g., body vitals monitoring including thermal patterns, as well as personal protective equipment (PPE) detection and PPE-wearing person identification. 
	
	A recent survey on Covid-19 pandemic \cite{[Alamo-2020]} covers {decision-making support} in many projections such as data-driving modeling, testing, tracing contacts, benchmarks, data hubs, machine learning, and privacy. 
	
	In \cite{[Fenton-2020]}, reasoning mechanism using a Bayesian causal network is used for studying \emph{collider bias} of Covid-19 disease risk and severity. In causal networks, a variable is a collider when it is causally influenced by two or more variables; this results in either over-estimation or under-estimation of causal effects. Paper \cite{[Fenton-2020]} also indicates potential Covid-19 collider biases caused by blood type, demographic factors, and related diseases. In \cite{[McLachlan-2020]}, the DSS concept is implemented as causal reasoning on contact tracing to reduce the Covid-19 spread by providing diagnostic-oriented feedback (user symptoms) to citizens with near real-time Covid-19 surveillance, as well as an accurate picture of the extent and patterns of disease transmission at the community-level for a constantly changing situation. Authors discuss various privacy risk mitigation approaches, public compliance, and trust. Causal reasoning upon the infection prevalence and fatality rates is used in \cite{[Neil-2020]}. 
	
	\subsubsection{DSS for autonomous systems} 
	
	According to the taxonomical view in \cite{[Danks_2017]}, the prevalent types of the \textbf{algorithmic biases} in autonomous systems include training data bias, algorithmic focus and processing biases, transfer context bias, and interpretation bias. Responses to the algorithmic bias, in particular, include identifying and intervening problematic biases. A related bias is known as the artificial intelligence bias \cite{[AI-now-Report-2018]}. Paper \cite{[deVisser-2018]} addresses the mitigation of the various kinds of biases in autonomous systems using the concept of human-machine \emph{trust repair}, described as a certain act that makes trust more positive. This is the same kind of action as a feedback loop in a cognitive DSS. 
	
	\subsubsection{Combat DSS} 
	
	Contemporary military combat teams include both soldiers and autonomous robots \cite{[Brown-2020]}. Situational awareness tasks for each soldier are supported in combat by a biometric-enabled, wearable DSS, e.g., stress and fatigue detector \cite{[Pandey-2016],[Schmidt-2019]}. This addresses the problem of individual effects of stress (cognitive, emotional, behavioral, and physiological) and team effect of stress such as decreased cooperation, ineffective communication, and decreased coordination. Decision-making in such a unit is radically different from a human-only team, since in a human-robot team, a portion of the responsibility is delegated to the intelligent machines. Rapid trust calibration becomes a task of high priority \cite{[Tomsett_2020]}. This problem formulation is known as a \emph{human-CI teaming situational awareness} \cite{hou2014intelligent}.
	
	\subsubsection{Ambient systems}
	
	Ambient adaptive systems such as ambient CI assistants or monitors of human occupant vitals/biometrics have to use mechanisms to regulate themselves and change their structure in order to operate efficiently within dynamic ubiquitous computing environments. As a consequence of the increasingly aging population, it is necessary to find solutions to improve the living condition and develop more robust, usable, safe, and low-cost healthcare systems. This leads to a fixed DSS to be incorporated in ambient systems such as smart home, mobility and health assistants \cite{[Li-2015]}. 
	
	\subsubsection{E-coaching} 
	
	E-coaching systems are aimed at supporting individuals in their self-regulation \cite{[Kamphorst-2017]} using various biometrics. E-coaching offers support in the following areas: social ability, credibility, context-awareness, personalization	(user tailoring), learning of user behavior, proactiveness, and guidance (coaching planning) \cite{[Kamphorst-2017],[Ochoa-2018]}. Measures of efficiency in such systems can naturally be expressed in terms of R-T-B. 
	
	\section{Conclusions}\label{sec:Summary_conclusions}
	
	Biometric-enabled systems are becoming an integral part of more complex intelligent systems. Such system-to-system embedding requires a deep unification of computational platforms and performance regulators. The proposed approach in this study use R-T-B as DSS performance indicators. These indicators shall become a mandatory assessment tool for all stages of the DSS development and deployment for the following reasons:
	\begin{enumerate}
		\item The R-T-B causal-based taxonomy provides efficient resources for deriving the knowledge (from biometrics) required for decision-making in biometric-enabled systems.
		\item The DSS core, the reasoning over the R-T-B projection is implemented using causal networks (including Bayesian); each of these network types provides specific interpretation and approximation of uncertainty stemmed from the nature of biometric data.
		\item The DSS with R-T-B indicators are most appropriate for forecasting applications, including risk assessment in the biometric-enabled systems that are planned to be implemented given the specific security, privacy, and usability scenarios. 			
	\end{enumerate}
	
	From a \emph{practical standpoint}, the R-T-B indicators are useful performance evaluation tools in any biometric-enabled system. From a \emph{theoretical standpoint}, measuring the R-T-B is an ultimate probabilistic and computational intelligent problem because it aims at the development of a proactive mechanism to detect ill-defined phenomena from observable data. The problem is extremely challenging because the R-T-B are conceptual constructs (often psychological indicators) that are not directly observable and are computed from the multiple sources of factors embedded in a noisy context within a system operation. Among various challenges, we emphasize that consensus methodology for the group DSS is an open problem (Fig. \ref{fig:Group-DSS}). 
	
	\section*{Acknowledgments}
	\begin{small}
		This Project was partially supported by the
		Natural Sciences and Engineering Research Council of Canada (NSERC) through grant ``Biometric-enabled Identity management and Risk Assessment for Smart Cities'', and by the Department of National Defence's Innovation for Defence Excellence and Security (IDEaS) program, Canada. The authors acknowledge Eur Ing \emph{Phil Phillips}, CEng (UK), for useful suggestions.
	\end{small}

\end{document}